\documentclass[pra,twocolumn,showpacs]{revtex4-1}
\usepackage{bm,amsmath,amsfonts,amsbsy}
\newcommand{\inn}[2]{\v{#1}\cdot\!\v{#2}}

\def \v#1{{\bm #1}}
\def \be {\begin{equation}}
\def \ee {\end{equation}}
\def \bml{\begin{multline}}
\def \eml{\end{multline}}
\newcommand{\Exp}[1]{\,\mathrm{e}^{\mbox{\footnotesize$#1$}}}
\newcommand{\I}{\mathrm{i}}

\newcommand{\ket}[1]{|#1\rangle}
\newcommand{\bra}[1]{\langle#1|}
\newcommand{\tr}[1]{\mathrm{tr}\left(#1\right)}

\newcommand{\expectn}[1]{\langle#1\rangle}
\def \ds{\displaystyle}

\def \del{\partial}

\def \cM{{\cal M}}
\def \cH{{\cal H}}
\def \cX{{\cal X}}
\def \sofh{{\cal S}({\cal H})}

\def \bbr{{\mathbb R}}
\def \bbc{{\mathbb C}}

\def \sofc2{{\cal S}({\mathbb C}^2)}
\def \sofhN{{\cal S}({\cal H}^{\otimes N})}

\def \cD#1{{\cal D}_{#1}}

\newtheorem{theorem}{Theorem}[section]
\newtheorem{lemma}[theorem]{Lemma}

\newtheorem{corollary}[theorem]{Corollary}

\newenvironment{proof}[1][Proof:]{\begin{trivlist}
\item[\hskip \labelsep {\bfseries #1}]}{\end{trivlist}}

\newcommand{\qed}{\nobreak \ifvmode \relax \else
      \ifdim\lastskip<1.5em \hskip-\lastskip
      \hskip1.5em plus0em minus0.5em \fi \nobreak
      \vrule height0.75em width0.5em depth0.25em\fi}
\begin{document}

\title{Entanglement detection from channel parameter estimation problem}
\author{Jun Suzuki\\
(Email: junsuzuki@is.uec.ac.jp)}  
%\email{junsuzuki@is.uec.ac.jp}
\date{\today}
\affiliation{Graduate School of Information Systems, 
The University of Electro-Communications,\\
1-5-1 Chofugaoka, Chofu-shi, Tokyo, 182-8585 Japan}

\begin{abstract}
We derive a general criterion to detect entangled states in multi-partite systems 
based on the symmetric logarithmic derivative quantum Fisher information. 
This criterion is a direct consequence of the fact that separable states 
do not improve the accuracy upon estimating one-parameter family of quantum channels.  
Our result is a generalization of the previously known criterion for 
one-parameter unitary channel to any one-parameter quantum channel. 
We discuss several examples to illustrate our criterion. 
The proposed criterion is extended to the case of open quantum systems and 
we briefly discuss how to detect entangled states in the presence of decoherence. 
\end{abstract}

%03.65.Wj: State reconstruction, quantum tomography 
%02.50.Tt; Inference methods 
%03.67.-a; Quantum information
%03.67.Mn; for entanglement measures, witnesses etc
%\keywords{Multi-partite entanglement, SLD quantum Fisher information, channel parameter estimation}

%########### body
\maketitle
\section{Introduction}
In this paper we address a problem of general relationship between 
entanglement in multi-partite systems and quantum Fisher information 
from channel-parameter estimation perspective. 
This problem has gained a great interest in the field of research so called quantum metrology, 
that is, quantum mechanically enhanced precision measurements \cite{glm11,review1,review2,review3}.  
Our work is motivated by the theoretical work of Pezz\'e and Smerzi and 
the experimental verification of their criterion \cite{ps09,smlzhpso14}. 
In Ref.~\cite{ps09}, they observed that the symmetric logarithmic derivative (SLD) 
quantum Fisher information 
for any separable state cannot be greater than the total number of qubits 
when qubit states undergo a global rotation along some axis. 
Their result was further generalized to detect $k$-producible states 
and to derive other criteria by taking averages with respect to rotation axes \cite{hps10,hlkswwps12,toth12,ll13}, 
see also updated references cited in review articles \cite{review1,review2,review3}.
In the recent paper \cite{smlzhpso14}, a beautiful experimental result has been reported 
showing that a global rotation of atomic spin-states was used to detect non-gaussian entangled states. 
The main objective of this paper is to generalize their criterion to any quantum channel. 

The second motivation of our work is to examine whether entanglement brings 
benefit upon estimating parameters for non-unitary channels, i.e., general quantum channels. 
To answer this question, we shall use a standard language of quantum parameter estimation theory 
developed by Helstrom, Yuen-Lax, Holevo, and others \cite{helstrom,yl73,holevo,hayashi,hayashi2}. 
A formal channel-parameter estimation problem in quantum systems was initiated by 
Fujiwara and his collaborators \cite{fujiwara01,fi03}, where they utilize basic tools developed 
in quantum parameter estimation theory mentioned above. 

There exist at least four known no-go theorems regarding the observation 
of quantum metrologically enhanced measurement upon estimating one-parameter family of quantum channels. 
Ji \textit{et al} \cite{jwdfy08} showed a rather remarkable result that 
any programmable channels cannot be estimated with 
quantum metrological enhancement. 
Here a given channel is programmable or not is defined  
by Ref.~\cite{nc97}. 
Fujiwara and Imai \cite{fi08} provided another no-go theorem 
stating that quantum metrological enhancement cannot occur 
for any full-rank channels changing smoothly with the parameter. 
Their result is very general and implies that almost all realistic quantum channels 
do not exhibit such quantum metrological enhancement. 
In the unpublished work, Matsumoto gave a simple criterion where 
any classically simulated channel cannot be estimated with quantum metrological enhancement \cite{matsumoto10}. 
Two results in Refs.~\cite{fi08,matsumoto10} are well summarized in the paper \cite{ddkm12}, 
where authors applied these two criteria for physically important quantum channels. 
Lastly, Hayashi \cite{hayashi11} gave a very powerful argument; 
no quantum metrological enhancement occurs when a given 
channel admits finite amount of the right logarithmic derivative (RLD) quantum Fisher information. 

These no-go theorems state that there are quantum channels 
in which we cannot utilize quantum entanglement 
to go beyond the standard quantum limit \cite{comment1}. 
This poses a question whether or not non-unitary 
channels which satisfy the above no-go criteria can be used 
to detect entanglement. 
In this paper, we answer this question first by showing that 
all separable states do not bring any benefit upon estimating 
one-parameter channels. This is then translated into 
a simple yet general criterion: 
If the amount of SLD quantum Fisher information of the output state 
for a given family of one-parameter quantum channel is above a certain threshold, 
then the input state mush be entangled. 
We then examine several examples to demonstrate the obtained criterion, 
such as the unitary channel, depolarizing channel, and transpose channel. 
We show that detection of entanglement is possible for certain range of parameter of these channels 
at least bi-partite case. 

This paper is organized as follows. 
In Sec.~\ref{sec2} we summarize notations and discuss relationship between 
classical information quantities and quantum Fisher information. 
In Sec.~\ref{sec3} we prove the main result of this paper. 
In Sec.~\ref{sec4} we apply our criterion to the bi-partite case and compare 
result for different quantum channels. 
In Sec.~\ref{sec5} we extend our criterion from the i.i.d. setting to a more general setting 
in order to apply it for open quantum systems. 
In the last section, we summarize our results. 

\section{Channel-parameter estimation in quantum systems}\label{sec2}
We provide basic terminologies and notations used in this paper.  
We then summarize the basic result known for one-parameter 
channel estimation problems in quantum systems. 

\subsection{Preliminaries}
Let $\cH$ be a finite dimensional Hilbert space 
and $\sofh$ be the set of all density operators on $\cH$, 
which are semi-definite positive. 
Let $\Gamma_\theta$ be a trace-preserving and completely-positive (TP-CP) map 
(also called a quantum channel, a quantum map, etc.)
from $\sofh$ to itself that is parametrized by a single 
parameter $\theta$: 
\be
\Gamma_\theta:\,\sofh\longrightarrow\sofh\ (\mathrm{TP-CP}).
\ee
Assume that the parameter $\theta$ takes 
values in an open subset of real numbers, $\Theta\subset{\mathbb R}$, then 
the output state $\rho_\theta=\Gamma_\theta(\rho)$ for a given input state $\rho\in\sofh$ 
can be regarded as a quantum statistical model parametrized this parameter $\theta\in\Theta$; 
\be \label{qmodel}
\cM=\left\{\rho_\theta=\Gamma_\theta(\rho)\, |\, \theta\in\Theta\subset{\mathbb R}\right\}.
\ee
Depending on the channel and the given input state, 
the rank of the output states $\Gamma_\theta(\rho)$ may vary with respect to the parameter $\theta$ in general. 
For mathematical convenience, we further assume that the rank of the quantum statistical model $\cM$ 
does not change for all values $\theta\in\Theta$ at least for each fixed input quantum state. 

The SLD operator about $\rho_\theta\in\cM$ is defined by 
an hermite operator $L_\theta$ satisfying the equation:
\be\label{sld}
\frac{d}{d\theta}\rho_\theta=\frac12 (\rho_\theta L_\theta+L_\theta\rho_\theta).
\ee 
The SLD quantum Fisher information about $\rho_\theta$ is defined by 
\be\label{sldFinfo}
g_\theta[\rho_\theta]:=\tr{\rho_\theta L_\theta^2} .
\ee   
By definition, it also holds that $g_\theta[\rho_\theta]=\tr{L_\theta \frac{d}{d\theta}\rho_\theta }$. 
For full-rank states, the solution to the above operator 
equation \eqref{sld} is unique, i.e., the SLD operator is uniquely defined. 
For low-rank states such as pure states, on the other hand, SLD operator is not uniquely determined 
from the above equation. 
In this case, one has to consider equivalent classes to define a proper inner product 
first and then to define SLD operator resulting in the unique SLD quantum Fisher information \cite{fn95}.  

There are several important properties of the SLD quantum Fisher information to be listed below. 
First, it is non-negative, i.e., $g_\theta[\rho_\theta]\ge 0$. 

Second, quantum Fisher information is additive, i.e., 
for any product state $\rho_\theta=\rho_\theta^1\otimes\rho_\theta^2\in{\cal S}(\cH_1\otimes\cH_2)$, 
\be
g_\theta[\rho_\theta^1\otimes\rho_\theta^2]=g_\theta[\rho_\theta^1]+g_\theta[\rho_\theta^2]. 
\ee

Third, it cannot increase when a CP-TP $\Gamma$ is applied to the state, i.e. 
the following inequality holds. 
\be
g_\theta[\Gamma (\rho_\theta)]\le g_\theta[\rho_\theta].  
\ee
This properly is usually referred to as the monotonicity of SLD quantum Fisher information \cite{petz}. 
As a special case of quantum channels, let us consider a general measurement, 
described by a positive operator-valued measure (POVM), 
$\Pi=\{\Pi_x\,|\,x\in\cX\}$. The map from a state $\rho$ to 
a probability distribution $p_\theta(x)=\tr{\rho_\theta E_x}$ 
is regarded as quantum to classical channel since 
the state $\rho_{\Pi}=\mathrm{diag}(p_\theta(x_1),p_\theta(x_2),\dots)$ 
describes the probability distribution for measurement outcomes. 
The SLD quantum Fisher information about this (classical) state is equal to the (classical) Fisher information 
$G_\theta^c[p_\theta]$. Since the probability distribution for measurement outcomes is 
determined by a given POVM $\Pi$, we also write it as $G_\theta^c[\Pi,\rho_\theta]$. 
Thus, the following inequality for any POVM holds. 
\be
g_\theta[\rho_\theta]\ge G_\theta^c [\Pi,\rho_\theta].
\ee
We call this property as the q-c (quantum to classical) monotonicity of the SLD quantum Fisher information. 

Last, it is convex with respect to quantum states. Let $\rho_\theta^j\in\sofh$ ($j=1,2$) be 
two families of states with the same parameter set $\Theta$. The convex property states 
\be\label{convineq}
g_\theta[\lambda\rho_\theta^1+(1-\lambda)\rho_\theta^2]\le \lambda g_\theta[\rho_\theta^1]+(1-\lambda)g_\theta[\rho_\theta^2],
\ee
for any $\lambda\in[0,1]$. This convexity can be proven by 
many ways. The simplest is given in Ref.~\cite{fujiwara01} using 
the monotonicity of the SLD quantum Fisher information. 
It seems that the equality condition for the above convex inequality is 
in general complicated. Since this condition is important, we examine 
it for a simple unitary model, which is given at the end of Sec.~\ref{sec3-2}. 

The main objective of channel-parameter estimation in quantum systems is 
to find the ultimate precision bound and the optimal strategy 
upon estimating the value of parameters describing a given channel. 
Here we stress that there is no unique way to define the optimality and 
one has to analyze a given problem according to a suitable figure of merit. 
A strategy upon estimating the value of the given quantum channel consists of three elements: 
An input state, a measurement, and an estimator. 
One way to get the optimal strategy is as follows. 
For a fixed input state $\rho$, we optimize over all possible quantum measurements described 
by a POVM $\Pi$ and an estimator $\hat{\theta}$ 
which is a classical data processing. The set $(\Pi,\hat{\theta})$ is called 
a quantum estimator or simply estimator in this paper. 
With this optimal estimator, we optimize over 
all possible input states available. A triplet $(\rho,\Pi,\hat{\theta})$ is called 
an estimation strategy for the quantum channel. 
For a one-parameter problem, this procedure 
gives at least asymptotically optimal one \cite{comment2}.  

When one concerns the mean-square error (MSE) as a figure of merit for the channel estimation, 
one can derive the lower bounds for the MSE depending upon resources and estimation schemes 
under consideration. 
Let  $\cH^{\otimes N}$ and $\sofhN$ be $N$ tensor product of Hilbert space and totality of positive density operators on it, respectively. 
Consider an $N$th i.i.d. extension of the given channel and denote it as 
\be
\Gamma_\theta^N:=\underbrace{\Gamma_\theta\otimes\Gamma_\theta\otimes\dots\otimes\Gamma_\theta}_{N}:\,\sofhN\longrightarrow\sofhN.
\ee
When one only uses $N$th i.i.d. extension of input states $\rho^{\otimes N}$ to estimate 
the channel, the problem is to find an optimal input state maximizing 
the SLD quantum Fisher information for the channel $\Gamma_\theta$. 
Let $\rho^*$ be one of such an optimal input state and $g_\theta^*$ be the maximum of the SLD quantum Fisher information;
\be\label{opt}
g_\theta^* (\Gamma_\theta):=\smash{\displaystyle\max_{\rho\in\sofh}} \big\{ g_\theta[\Gamma_\theta(\rho)] \big\}.
\ee
Importantly, the convexity property of the SLD Fisher information 
guarantees that the optimal input state attaining $g_\theta^* (\Gamma_\theta)$ 
can be a pure state \cite{fujiwara01}. 

The additivity of SLD quantum Fisher information concludes 
$g_\theta (\Gamma_\theta^N(\rho^{\otimes N}))=Ng_\theta (\Gamma_\theta(\rho))$. 
For any locally unbiased estimators $(\Pi,\hat{\theta})$, 
the MSE is equal to the variance of estimating the value of parameter and is bounded  as
\be
\mathrm{Var}_\theta[\Pi,\hat{\theta}]\ge \frac{1}{N} \left(g_\theta^* (\Gamma_\theta)\right)^{-1}.
\ee
In general, this bound is attained adaptively in the $N$ infinite limit unless the channel possesses 
a special symmetry.  
See the discussion given in Ref.~\cite{fujiwara06} and 
an experimental demonstration of the adaptive estimation \cite{oioyift12}. 
Alternatively, one can use the two-step method proposed in Refs.~\cite{HM98,BNG00}. 

When one estimates the $N$th i.i.d. extension of the channel $\Gamma_\theta^N$, 
one can also use other resources such as entangled states $\rho\in\sofhN$ for input states or ancillary states. 
In this case, the variance for estimation can be further lowered. This enhancement effect, 
known as quantum metrology, is of importance for quantum information processing 
protocols and has been investigated actively \cite{glm11,review1,review2}. 

\subsection{Experimental detection of SLD quantum Fisher information}
In this subsection, we discuss a general strategy how to detect the amount of 
quantum Fisher information about the output state of a given family of quantum channels in experiment. 
We assume that the parameter for the quantum channel can be tuned at will 
and there are identical resources to repeat the same experiment sufficiently many times. 
A prominent step was already reported in Ref.~\cite{smlzhpso14}. In this paper we 
shall present a more general framework to supplement their result. 

For a given one-parameter family of quantum channels $\Gamma_\theta$, let us 
fix the input state $\rho$ and consider a fixed measurement $\Pi$ on the 
output state $\rho_\theta=\Gamma_\theta(\rho)$. Then, 
the family of probability distributions for the measurement outcomes is 
regarded as a classical statistical model: 
\begin{align}
\cM(\Pi,\Gamma_\theta,\rho)&=\{p_\theta[\Pi] \,|\, \theta\in\Theta\},\\ \nonumber 
p_\theta[\Pi]&=\{ p_\theta(x)=\tr{\Gamma_\theta(\rho)\Pi_x}\,|\,x\in\cX \}. 
\end{align}
By performing sufficiently large repetition of the same measurement for a fixed 
value of the parameter $\theta$, we can obtain 
experimental data according to the classical probability distribution $p_\theta[\Pi]$.  
We next change the channel parameter $\theta$ and redo the 
same step as before. After sufficiently many observations with respect 
the changes in $\theta$, say $M$ different choices, 
we can obtain the set of classical probability distributions 
$\{p_\theta\,|\,\theta\in \{\theta_1,\theta_2,\dots, \theta_M \}\}$.
If we choose the parameter set $ \{\theta_1,\theta_2,\dots, \theta_M \}$ 
($\theta_{k+1}>\theta_k$) such that the differences $\Delta_k=\theta_{k+1}-\theta_{k}$ 
is sufficiently small, then one can directly calculate the classical Fisher information 
$G^c_\theta[p_\theta]$ approximately from the definition: 
\be
G^c_\theta[p_\theta]:= \sum_{x\in\cX} \frac{[\frac{d}{d\theta} p_\theta(x)]^2}{p_\theta(x)}. 
\ee

Alternatively, one can estimate other information quantities first 
and then to calculate the classical Fisher information as follows. 
In classical information theory, the general information quantity is 
the f-divergence \cite{csbook}. This family of information quantity 
is a measure of ``distance'' between two probability distributions. 
The formal definition of f-divergence for two probability distributions $p,q$ on $\cX$ is 
\be
D_f(p||q):= \sum_{x\in\cX} p(x) f\Big(\frac{q(x)}{p(x)}\Big), 
\ee
where $f: \bbr^+\to\bbr^+$ is a monotonically decreasing and convex function 
and $f(1)=0$ is a standard convention. 
Familiar examples are: $f(t)=-\log(t)$ (the relative entropy), $f(t)=1-\sqrt{t}$ (the Hellinger distance), $f(t)=t^\alpha$ (the relative R\'enyi entropy).
One of important properties of the f-divergence is that 
the following relation to Fisher information: 
\be \label{ftofisher}
G^c_\theta[p_\theta]=2\lim_{\epsilon\to0}\ \frac{1}{\epsilon^2} D_f(p_\theta||p_{\theta+\epsilon}). 
\ee
From experimental data obtained after many repetitions, we can construct a curve for f-divergence 
$D_f(p_\theta||p_{\theta'})$ for various different values of $\theta,\theta'$. 
It is easy to see that the formula \eqref{ftofisher} provides 
an approximated value for the classical Fisher information. 

We next show that this experimentally obtained Fisher information can 
attain the SLD quantum Fisher information by the optimal measurement. 
As remarked in the previous subsection, q-c monotonicity of 
the SLD quantum Fisher information implies 
\be
g_\theta[\Gamma_\theta(\rho)]\ge G^c_\theta[\Pi], 
\ee
where the equality holds if and only if $\Pi$ is optimal one 
and it is given by the projection measurement about the SLD operator $L_{\theta}$ \cite{yang,nagaoka87,bc94}. 
By choosing the optimal measurement, the classical Fisher information obtained from the above described method 
yields the approximated value of the quantum Fisher information.

\section{Result}\label{sec3}
\subsection{Separability criterion}\label{sec3-1}
The main result of this paper is the following theorem:
\begin{theorem}\label{thm1}
For a given channel $\Gamma_\theta$ parametrized by a single parameter $\theta$, 
let $\Gamma_\theta^N$ be an $N$th i.i.d. extension of $\Gamma_\theta$ and $g_\theta^*(\Gamma_\theta)$ be the largest value 
of SLD quantum Fisher information, which is given by Eq.~\eqref{opt}.  For each value $\theta$, 
if a density operator $\rho$ on $\sofhN$ is separable, then the SLD quantum Fisher information 
$g_\theta[\Gamma_\theta^N(\rho)]$ is smaller or equal to the value $N g_\theta^*(\Gamma_\theta) $. 
\end{theorem}

Several remarks are in order. 
First, taking the contraposition of this theorem, 
it is equivalent to state if the value of SLD quantum Fisher information for the output states $\Gamma_\theta^N(\rho)$ 
is larger than $N g_\theta^* (\Gamma_\theta)$, then the input state $\rho$ on $\sofhN$ is entangled. 
Second, the special case of this separability criterion was stated by Pezz\'e and Smerzi \cite{ps09}, 
where the channel is given by a rotation along a given axis on $N$ qubits in the context of quantum metrology.  
In this case, $g_\theta^* =1$ holds for all values of $\theta$ due to symmetry of a shift-parameter model. 
This special case will be examined in the next subsection. 
Thus, our contribution is to first prove their criterion in a general setting 
and to provide more general criterion. 

Third, since the MSE for estimation of the value $\theta$ is bounded 
by the inverse of SLD quantum Fisher information, Theorem \ref{thm1} states that separable states 
are not efficient upon the usage of $N$th extension of a given channel. 
But, of course, this theorem does not tell if all entangled states are more efficient than 
separable ones or not. 

Fourth, it is straightforward to see this theorem can be extended to more general channels 
parameterized by several parameters. In this case, the SLD quantum Fisher information becomes 
a matrix and the corresponding inequality is given by a matrix inequality. 
It is also not difficult to see from the proof that SLD quantum Fisher information can 
be replaced by other quantum Fisher information. 

Lastly, the most important is that the parameter $\theta$ is arbitrary value in Theorem \ref{thm1}. 
Since we can vary it as an arbitrary value, 
one take a union of all possible parameter regions of entangled states.  
Let $r_{ent}(\theta)$ be the entangled region of the states derived from 
the inequality $g_\theta[\Gamma_\theta^N(\rho)]> N g_\theta^*(\Gamma_\theta) $, i.e., 
\be \label{loregion}
r_{ent}(\theta)=\left\{\rho\in\sofhN \,|\, g_\theta[\Gamma_\theta^N(\rho)]> N g_\theta^*(\Gamma_\theta)\right\}, 
\ee
then, the union 
\be \label{region}
R_{ent}:=\bigcup_{\theta\in\Theta} r_{ent}(\theta),
\ee
provides the most powerful criterion. Since the subset of states $R_{ent}\subset\sofhN$ 
is solely determined by the given quantum channel $\Gamma_\theta$, we denote it as 
$\rho\in R_{ent}[\Gamma_\theta]$. 
With these notations, our contribution is to derive the new criterion:
\be\label{entcrit}
\rho\in R_{ent}[\Gamma_\theta]\ \Rightarrow\ \rho\mbox{ is entangled}. 
\ee
This point will be illustrated by several examples in Sec.~\ref{sec4}. 

Proof for Theorem \ref{thm1} is straightforward and is given as follows:
\begin{proof}
Consider an arbitrary separable states on $\sofhN$ of the form
\be
\rho_{sep}=\sum_{j}p_j\rho_j^{(1)}\otimes\rho_j^{(2)}\otimes \dots\otimes\rho_j^{(N)},
\ee
where $\sum_jp_j=1,\forall p_j\ge0$ and $\rho_j^{(k)}$ are states on the $k$th Hilbert space.  
Then, the following sequence of inequalities holds.
\begin{align}
g_\theta[\Gamma_\theta^N(\rho_{sep})]&=
g_\theta[\sum_jp_j\Gamma_\theta^N(\rho_j^{(1)}\otimes \dots\otimes\rho_j^{(N)})] \\ \label{conv}
&\le \sum_{j}p_j g_\theta[\Gamma_\theta^N(\rho_j^{(1)}\otimes \dots\otimes\rho_j^{(N)})]\\
&=\sum_{j}p_j g_\theta[\Gamma_\theta(\rho_j^{(1)})\otimes \dots\otimes\Gamma_\theta(\rho_j^{(N)})]\\
&=\sum_{j}p_j \sum_{k=1}^{N}g_\theta[\Gamma_\theta(\rho_j^{(k)})]\\ \label{defofg}
&\le\sum_{j}p_j \sum_{k=1}^{N}g^*_\theta[\Gamma_\theta]\\
&=Ng^*_\theta[\Gamma_\theta].
\end{align}
The inequality in \eqref{conv} follows from the convexity of SLD quantum Fisher information 
with respect to states and the inequality \eqref{defofg} does from the definition of $g^*_\theta[\Gamma_\theta]$. 
$\square$
\end{proof}

\subsection{Shift-parameter model}\label{sec3-2}
As noted in the remarks, the above theorem is simplified when the channel 
is given by a unitary transformation of the form: 
\be
\Gamma_\theta(\rho)=\Exp{\I \theta A} \rho\Exp{-\I \theta A},
\ee
where an hermite operator $A$ on $\cH$ is called a generator of the unitary transformation. 
The parameter region is any $2\pi$ interval of real numbers, e.g., $\Theta=[0,2\pi)$. 
The quantum statistical model about the output states is given by 
\be \label{spmodel}
\cM_A=\left\{\rho_\theta=\Exp{\I \theta A} \rho_0\Exp{-\I \theta A}\, |\, \theta\in{\mathbb R}\right\}.
\ee
Here, $\rho_0$ is called as a reference state. 
This model was referred to as a shift-parameter model or a displacement model in Refs.~\cite{helstrom,holevo}. 

The following lemma is fundamental for the unitary model. 
\begin{lemma} \label{lem1}
For a shift-parameter model, the SLD quantum Fisher information is independent of the parameter $\theta$ 
and is bounded from above as
\be \label{eq:lem1}
g_\theta=g_{\theta=0}\le4\Delta _{\rho_0}A, 
\ee
where $\Delta _{\rho} A:=\tr{\rho A^2}-(\tr{\rho A})^2$ is the square of variance of the 
operator $A$ with respect to the state $\rho$. 
\end{lemma}
This lemma can be proven in different manners, here we sketch the most transparent one 
due to Holevo \cite{holevo}. 
\begin{proof}
For a given state $\rho$, let $\cD{\rho}$ be a super-operator acting on hermite operators $X$ on $\cH$, 
which is formally defined by the solution to the following operator equation;
\be
\rho\, \cD{\rho}(X)+\cD{\rho}(X)\, \rho=\frac{1}{\I} [\rho\,,\,X]. 
\ee
It follows from the definition that the SLD operator is expressed as 
\begin{align}
L_\theta&=2\cD{\rho_\theta}(A)=\Exp{-\I \theta A}L_0\Exp{\I \theta A},\\
L_0&= 2\cD{\rho_0}(A).
\end{align}
This relation proves the first equality in Eq.~\eqref{eq:lem1}. 

We define a symmetric inner product for linear operators $X,Y$ on $\cH$ by
\be
\expectn{X,Y}_\rho:=\frac12 \tr{\rho(YX{^\dagger}+X^{\dagger}Y)}, 
\ee
then the SLD quantum Fisher for the shift-parameter model \eqref{spmodel} is written as 
\begin{align} \nonumber
g_0&= \expectn{L_0,L_0}_{\rho_0}\\ 
&=4 \expectn{\cD{\rho_0}(A),\cD{\rho_0}(A)}_{\rho_0}. 
\end{align}
Next, we note that the relation 
\be 
\expectn{X,X}_\rho-\expectn{\cD{\rho}(X),\cD{\rho}(X)}_\rho=\expectn{X,(1+\cD{\rho}^2)(X)}_\rho\ge0,
\ee
holds for any hermite operators $X$ and a state $\rho$ on $\cH$, 
since the super-operator $1+\cD{\rho}^2$ is positive with respect to the inner product. 
Writing the variance as $\Delta _{\rho_0}A=\expectn{A-\bar{A},A-\bar{A}}_{\rho_0}$ 
with $\bar{A}=\tr{\rho A}$ and using the relation $\cD{\rho_0}(A)=\cD{\rho_0}(A-\bar{A})$,  
we prove the inequality in Eq.~\eqref{eq:lem1}.
$\square$
\end{proof}

We note that equality condition for the inequality in Eq.~\eqref{eq:lem1} 
is equivalent to the condition \cite{holevo}:
\be \label{hcond}
(1+\cD{\rho}^2)(A-\bar{A})=0\Leftrightarrow \rho_0A\rho_0=\bar{A}\rho_0^2.
\ee 
In some literature it is stated that ``the equality in Eq.~\eqref{eq:lem1} 
is satisfied if and only if $\rho_0$ is a pure state." 
This is true if the dimension of Hilbert space is $2$, i.e., qubit. 
However, we remark that the condition $\rho_0$ is pure is just a sufficient condition in general. 
The sufficiency is immediate if we uses the second condition in Eq.~\eqref{hcond}. 
A simple counter example of mixed states satisfying the upper bound is given by 
a rank $2$ state in dim $\cH$=3 as follows. 
\be
\rho_0=\left(\begin{array}{ccc}\lambda & 0 & 0 \\0 & 1-\lambda & 0 \\0 & 0 & 0\end{array}\right),\quad
A=\left(\begin{array}{ccc}a & 0 & c^* \\0 & a & d^* \\c & d & b\end{array}\right), 
\ee
where $\lambda\in(0,1)$, and $a,b$ and $c,d$ are real and complex numbers, respectively. 

The variance of all possible states $\rho\in\sofh$ is maximized when we take 
an equal weighted superposition of the eigenstates whose eigenvalues are maximum and minimum \cite{glm06}. 
With this observation, theorem \ref{thm1} and lemma \ref{lem1} can be combined to give the following corollary: 
\begin{corollary}
For a given shift-parameter model, let $U_A^N=\bigotimes_{k=1}^{N}\Exp{\I \theta A}$ 
be a global unitary for $\sofhN$. 
If a density operator $\rho$ on $\sofhN$ is separable, then the SLD quantum Fisher information 
$g_\theta[U_A^N\,\rho\, (U_A^N)^\dagger]$ is smaller or equal to the value $N (a_{max}-a_{min})^2$, 
where $a_{max}$ ($a_{min}$) is the maximum (minimum) of the eigenvalues of $A$. 
\end{corollary}

Since we have an achievable bound for unitary channels, one 
can consider various extensions of the criterion. An immediate one 
is to consider a set of rotations around certain axises and to take 
average SLD  Fisher information. Another one is 
to consider $k$-producible states rather than completely separable state of $N$ qubits. 
These extensions seem to work quite well as reported in Refs.~\cite{hlkswwps12, toth12, ll13}. 

Before moving to the next section to apply our criterion to examples, 
we shall analyze the equality condition for the convexity of 
SLD quantum Fisher information, i.e., the equality condition in the inequality \eqref{convineq}, 
in the case of this simple unitary model. 
Consider mixed qubit states generated by the following unitary: 
\be
\rho_\theta^{j}=\Exp{-\I\theta\inn{n}{\sigma}/2}\rho_{0}^{j}\Exp{\I\theta \inn{n}{\sigma}/2} \quad(j=1,2),  
\ee
where $\v n$ is a given unit vector and $\theta$ is the rotation angle 
and their convex mixture
\be
\rho_{\theta}^\lambda=\lambda\rho_\theta^1+(1-\lambda)\rho_\theta^2.
\ee 
Let $g_{\theta}^j=g_\theta[\rho_\theta^j]$ and $g_\theta^\lambda=g_\theta[\rho_\theta^\lambda]$ 
be the SLD quantum Fisher information about the state $\rho_\theta^{j}$ and $\rho_\theta^\lambda$, respectively. 
A straightforward calculation shows that 
\be
g_{\theta}^j=g_{\theta=0}^j= |\v n\times \v{s}_j|^2,
\ee
when is expressed in terms of the Bloch vector of the state; $\rho_0^j=(I+\v{s}_j\cdot\v{\sigma})/2$. 
Let us define the difference $\Delta g_\theta^\lambda=\lambda g_{\theta}^1+(1-\lambda)g_{\theta}^2-g_{\theta}^\lambda$, 
then it reads 
\be
\Delta g_\theta^\lambda=\lambda(1-\lambda)\big|\v{n}\times (\v{s}_1-\v{s}_2) \big|^2.
\ee
Therefore, the equality in the convexity inequality \eqref{convineq} holds 
if and only if the difference of the two Bloch vectors $\v{s}_1-\v{s}_2$ is 
parallel to the rotation direction $\v n$. This is 
equivalent to satisfying the condition $\v{s}_2=\v{s}_1- 2(\v{n}\cdot\v{s}_1)\v{n}$.  

\section{Examples}\label{sec4}
In this section we analyze several examples to illustrate the proposed criterion 
to detect entanglement, in particular, the criterion \eqref{entcrit}. 
To get analytical results, we simplify the setting to the two-qubit case, that is $N=2$ and dim $\cH$=2. 
The input states analyzed in this section are 
the Bell-diagonal states defined by 
\be
\rho_{BD}(c_1,c_2,c_3)=\frac14\Big(I+\sum_{j=1}^3c_j\sigma_j\otimes\sigma_j  \Big).
\ee
Here $\sigma_j$ are usual Pauli spin operators and the coefficients are restricted from the positivity condition as 
\begin{align} \nonumber
1-c_1-c_2-c_3&\ge0,\\\nonumber
 1-c_1+c_2+c_3&\ge0,\\\nonumber
1+c_1-c_2+c_3&\ge0,\\ 
1+c_1+c_2-c_3&\ge0. 
\end{align}
Among the Bell-diagonal states, we focus on the two sub-familes, 
\begin{align}
\rho_{\lambda}^+&:=\rho_{BD}(\lambda,\lambda,-\lambda),\\
\rho_{\lambda}^-&:=\rho_{BD}(-\lambda,-\lambda,-\lambda).
\end{align}
The states $\rho_{\lambda}^{\pm}$ are also written as 
\be
\rho_{\lambda}^{\pm}:=\lambda\ket{\psi_\pm}\bra{\psi_\pm}+\frac14(1-\lambda)I,
\ee
with $\ket{\psi_\pm}=(\ket{01}\pm\ket{10})/\sqrt{2}$ are the Bell states. 
For $\lambda$ in $\Lambda:=(-1/3,1)$, both states $\rho_{\lambda}^{\pm}$ 
are strictly positive. Further, $\rho_{\lambda}^{\pm}$ are entangled 
if and only if $\lambda\in \Lambda_{ent}:=(1/3,1]$. 
The difference between $\rho_{\lambda}^{\pm}$ is that 
$\rho_{\lambda}^{-}$ is rotationally invariant state (spin singlet state), 
whereas $\rho_{\lambda}^{+}$ is only invariant around the $z$-axis.

Our concern is to find a set of entangled states which can be detected 
by a given quantum channel. This quantity is represented by 
Eq.~\eqref{loregion} or Eq.~\eqref{region}. 
For the above family of states, states are uniquely 
specified by a parameter $\lambda$, detected entangled regions 
are expressed by some interval which is a subset of $ \Lambda_{ent}$.  

%%%%%%%%%%%%%%% 
\subsection{Unitary channel}
We first consider a rotation around the $z$-axis on a single qubit system as 
\be
U_z(\theta)=\Exp{\I\theta \sigma_z/2}, 
\ee
with $\theta\in\Theta$. The maximum variance of the generator $\sigma_z/2$is $1$, 
and Theorem \ref{thm1} reduces to the Pezz\'e-Smerzi criterion as discussed before. 
It compare the value of SLD quantum Fisher information 
about the state $U_z^N(\theta) \rho (U_z^N(\theta))^\dagger$ with the 
total number of qubit systems, i.e., $N$. 
Here, $U_z^N(\theta)=\bigotimes_{k=1}^{N}\Exp{\I \theta \sigma_z/2}=\Exp{\I\theta J_z}$ with $J_z$ 
the $z$ component of the total angular momentum operator.
As noted before, the SLD quantum Fisher information for the state $\rho_{\lambda}^{\pm}$ is zero 
due to the fact that $J_z$ commutes with $\rho_{\lambda}^{\pm}$. Thus, one 
cannot get any useful information about $\rho_{\lambda}^{\pm}$ by applying 
any global rotation around the $z$-axis. 

We next consider a rotation around the $x$-axis as  
\be
U_x(\theta)=\Exp{\I\theta \sigma_x/2}. 
\ee
The SLD quantum Fisher information about $\rho_{\lambda}^{+}$ is calculated as
\be
g_\theta[U_x^2\,\rho\, (U_x^2)^\dagger] =\frac{8\lambda^2}{1+\lambda} . 
\ee
Since the maximum SLD quantum Fisher information for the single system is $1$ 
as discussed in Sec.~\ref{sec3-2}, the Pezz\'e-Smerzi criterion states 
the state $\rho_{\lambda}^{+}$ is entangled if the following inequality holds. 
\be
\frac{8\lambda^2}{1+\lambda}> 2\ \Leftrightarrow\ 8\lambda^2-2\lambda-2>0.
\ee
Solving this inequality leads to the sufficient condition for the entangled region:
\be
R_{ent}=\Big(\frac{\sqrt{17}+1}{8},1\Big).  
\ee
The numerical value $(\sqrt{17}+1)/8\simeq0.64$ is larger than the true boundary $1/3$ as it should be. 

This example shows that entanglement in the state $\rho_{\lambda}^{-}$
cannot be detected by a rotation around any axis. 
For the state $\rho_{\lambda}^{+}$, on the other hand, 
a rotation around the $x$-axis can detect entanglement.

%%%%%%%%%%%%%%% 
\subsection{Depolarizing channel}
The depolarizing channel for a two-dimensional quantum system is defined by 
\be
\Gamma_\theta (\rho):=\theta\rho+\frac{1-\theta}{2} \tr{\rho} I. 
\ee
Here, the channel parameter $\theta$ represents a probability of errors taking values 
in $\Theta=(0,1)$, e.g., no error $\Leftrightarrow$ $\theta=1$ \cite{comment3}. 

Optimal parameter estimation strategies for this channel were studied based on 
various figures of merit, for example Refs.~\cite{fujiwara01,sbb02}. 
It was shown that this channel is programmable and hence $\theta$ 
cannot be estimated with quantum metrological enhancement \cite{jwdfy08}. 
The maximum value of the SLD quantum Fisher information for a single input state is given by 
an arbitrary pure state as
\be
g_\theta^*[\Gamma_\theta]=\frac{1}{1-\theta^2}.
\ee
The SLD quantum Fisher information are same for the two input states $\rho_{\lambda}^{\pm}$ 
and are calculated as
\be
g_\theta[\Gamma_\theta^2(\rho_{\lambda}^{\pm})]=\frac{12\theta^2\lambda^2}{(1-\theta^2\lambda)(1+3\theta^2\lambda)}. 
\ee
Thus, a sufficient condition for entanglement obtained from Theorem \ref{thm1} is 
$g_\theta[\Gamma_\theta^2(\rho_{\lambda}^{\pm})]>2g_\theta^*[\Gamma_\theta]$, 
equivalently, 
\be
3\theta^2(2-\theta^2)\lambda^2-2\theta^2\lambda-1>0. 
\ee
This inequality then gives the entangled region for $\lambda$ as 
\be \label{entregiondpc}
r_{ent}(\theta)=\Big(\frac{\theta+\sqrt{3(2-\theta^2)}}{3\theta (2-\theta^2)},1\Big),
\ee
which depends explicitly on the value of the channel parameter $\theta$. 
An important remark is that the parameter $\theta$ needs to satisfy 
$\theta\in[ \theta_c, 1]$ in order for the depolarizing channel to detect entanglement successfully. 
Otherwise, the criterion cannot tell if the states are entangled or not. Here 
the threshold is found numerically as $\theta_c\simeq 0.551$. Physically speaking, 
the channel cannot be too noisy to detect entangled states. 

Since this sufficient condition holds for any $\theta\in(\theta_c,1)$, 
the most useful one is given by the union as
\be
R_{ent}=\bigcup_{\theta\in\Theta}r_{ent}(\theta)=( \lambda_*,1), 
\ee
where $\lambda_{DPC}$ is the minimum of the function appearing in the expression \eqref{entregiondpc}, 
and is given by
\be
\lambda_{DPC}:=\min_{\theta\in\Theta}\frac{\theta+\sqrt{3(2-\theta^2)}}{3\theta (2-\theta^2)}
=1+\omega-\omega^{-1},
\ee
with $\omega=(\sqrt{2}-1)^{1/3}/2$. 
The numerical value $\lambda_{DPC}\simeq 0.837$ is larger than the one from a rotation around the $x$-axis. 

We note that Ref.~\cite{bff01} analyzed a parameter estimation problem of the depolarizing 
channel based on a specific measurement and an estimator. They observed that 
entangled states are superior to separable states for a certain sub-family of the Werner state. 
The numerical value found in Ref.~\cite{bff01} is close to the value reported in this paper, 
yet they are different by nature of problem. 

%%%%%%%%%%%%%%% 
\subsection{Transpose channel}
In this last example, we shall analyze a rather unusual channel defined in terms of 
a transpose operation. It is known that transposition operations are not 
completely positive, but only $1$-positive. Here, a key point is that the convexity of the SLD quantum Fisher 
information follows from monotonicity of SLD quantum Fisher information about arbitrary $1$-positive maps. 
Thus, trace-preserving and $1$-positive map is also capable of detecting entanglement. 

We consider the following channel from $\sofc2$ to itself:
\be
\Gamma_\theta(\rho):= \theta \rho+(1-\theta)\rho^T,
\ee
with $T$ the transpose operation. Here the parameter $\theta$ takes values in $\Theta=(0,1)$. 
The maximum value of the SLD quantum Fisher information when one uses a single qubit 
input state is 
\be
g_\theta^*[\Gamma_\theta]=\frac{1}{\theta(1-\theta)},
\ee
which is attained by the eigenstates of $\sigma_y$. 
A straightforward calculation gives the same values of the SLD quantum Fisher information 
for the input states $\rho_{\lambda}^{\pm}$ and is given by 
\be
g_\theta[\Gamma_\theta^2(\rho_{\lambda}^{\pm})]=8f_\theta\lambda^2
\left[\frac{1}{1-f_\theta\lambda}+\frac{1+f_\theta\lambda}{(1+f_\theta+2\lambda)^2-4\lambda^2}
\right],
\ee
with $f_\theta=(1-2\theta)^2$. 

A sufficient condition for entanglement is provided by Theorem \ref{thm1}: 
$g_\theta[\Gamma_\theta^2(\rho_{\lambda}^{\pm})]>2g_\theta^*[\Gamma_\theta]$.  
This is equivalently expressed as
\begin{multline} \label{tpccond}
F_\theta(\lambda):=4f_\theta(1-f_\theta)\lambda^4+f_\theta(f_\theta^2-2f_\theta+4)\lambda^3\\
+(f_\theta^2-2f_\theta-4)\lambda^2+f_\theta\lambda+1<0. 
\end{multline}
Detail analysis on this quartic equation $F_\theta(\lambda)=0$ shows that 
there are four real roots for all value of $\theta\in\Theta$. 
The relevant entangled region is then found as 
\be
r_{ent}(\theta)=\Big(\lambda_2(\theta),\frac{1}{2-f_\theta}\Big).
\ee
Here, $\lambda_2(\theta)$ is the second largest solution to the quartic equation $F_\theta(\lambda)=0$. 
Numerically, $\lambda_2(\theta)$ varies from $1/2$ to $1$ depending on the value of $\theta$. 
As in the depolarizing channel, we take union of $r_{ent}(\theta)$ to get the most useful criterion:
\be
R_{ent}=\bigcup_{\theta\in\Theta}r_{ent}(\theta)=\Big(\lambda_{TPC},1\Big), 
\ee
where $\lambda_{TPC}=\min_{\theta\in\Theta}\lambda_2(\theta)=1/2$. 

\subsection{Comparison and discussion}
In this last section, we compare four different channels studied in the previous sections 
and discuss our result. 
\begin{align*}
\mbox{Rotation around $z$-axis; }&U_z(\theta)=\Exp{\I\theta \sigma_z/2},\\ 
\mbox{Rotation around $x$-axis; }&U_x(\theta)=\Exp{\I\theta \sigma_x/2},\\  
\mbox{Depolarizing channel (DPC); }&\Gamma_\theta (\rho):=\theta\rho+\frac{1-\theta}{2} \tr{\rho} I,\\ 
\mbox{Transpose channel (TPC); }&\Gamma_\theta(\rho):= \theta \rho+(1-\theta)\rho^T.
\end{align*}
The result for $\rho_{\lambda}^{+}$ are summarized in Table 1. 
Results for $\rho_{\lambda}^{-}$ are same for the depolarizing and transpose channels. 
In Table 1, ``No" is indicated if a channel cannot be used to detect entangled states. 
$\lambda_2(\theta)$ is the second largest solution to the quartic equation $F_\theta(\lambda)=0$ (Eq.~\eqref{tpccond}). 
Numerically, $\lambda_2(\theta)$ varies from $1/2$ to $1$ depends on the value of $\theta$. 
\renewcommand{\arraystretch}{1.2} 
\begin{table}[htdp]
\begin{center}
\begin{tabular}{c||c|c|c}
&$\theta$ dependence&$r_{ent}(\theta)$&$\displaystyle R_{ent}$\\[0ex]
\hline
$U_z$&No&No&No\\[1ex]
\hline
$U_x$&No&$\ds(\frac{\sqrt{17}+1}{8},1)$&$\ds(\frac{\sqrt{17}+1}{8},1)$\\[1ex]
\hline
DPC&Yes&$\ds(\frac{\theta+\sqrt{3(2-\theta^2)}}{3\theta (2-\theta^2)},1)$&$\ds( \lambda_{DPC},1)$\\[1ex]
\hline
TPC&Yes&$\ds(\lambda_2(\theta),\frac{1}{2-(1-2\theta)^2})$&$\ds(\frac12,1)$
\end{tabular}
\end{center}
\label{default}
\caption{Summary of entanglement detection for the state $\rho^+(\lambda)$ from four different channels; 
two unitary channels, depolarizing channel (DPC), and transpose channel (TPC). 
Entanglement region $r_{ent}(\theta)$ is defined in Eq.~\eqref{loregion} and 
its union is denoted by $R_{ent}$ defined by Eq.~\eqref{region}. 
Numerical values are $ \lambda_{DPC}\simeq 0.837$ $>$ $(\sqrt{17}+1)/8\simeq0.64$ $>1/2$.}
\end{table}%

As noted before, the rotation around any axis is not useful for 
the Werner state $ \rho_{\lambda}^{-}$, since SLD quantum Fisher information 
about the output states is always zero. 
As we can see from Table 1, for $\rho_{\lambda}^{+}$, 
the rotation around the $x$-axis can be used to detect entanglement which 
performs better than the depolarizing channel. Interestingly, the 
(unphysical) transpose channel can detect entangled states 
better than other examples analyzed in this paper. 

The main difference between unitary channels and non-unitary channels is 
that SLD quantum Fisher information is $\theta$ independent for the unitary case. 
This might be an advantage in realistic situation if one wishes to detect entangled states 
with unknown unitary channel. For our point of view, however, this is not a problem, 
since we are willing to detect entangled states by engineering appropriate quantum channels. 

Experimentally, we prepare a family of quantum channels $\Gamma_\theta$ with 
a controlable parameter $\theta$ . 
We next send an unknown multi-partite state and perform a good measurement on the output state. 
The measurement results then give probability distributions depending on the value of the parameter $\theta$. 
We can then calculate classical Fisher information, which coincides with the SLD quantum Fisher 
information if the measurement is chosen as the optimal one. By comparing the 
value of Fisher information for multi-partite states with the optimal Fisher 
information for single input state, which is exploit in advance, one 
can tell if the states are entangled or not based on the criterion given in Theorem \ref{thm1}. 

Lastly, we show that for a certain parameter range (low-noise regime), the depolarizing 
cannel can be estimated more efficiently if we use entangled input states. 
Although we cannot get a full benefit from entanglement to attain quantum metrological 
enhancement, entanglement indeed enables the estimation error lower than the separable input states. 
Whether this effect is significantly important depends on how accurate one wishes 
to estimate the value of a parameter of a given channel. More analysis on other quantum channels 
as well as various entangled input states are needed to make any general statement. 

\section{Extension to open quantum system}\label{sec5}
So far we have concerned with the i.i.d. extension of quantum channels only.  
In this section, we shall extend the proposed criterion for the non i.i.d. case 
and discuss how to apply it to open quantum systems briefly. 
Let $\Gamma_\theta$ be a quantum channel 
from quantum states on $\cH^N:=\cH_1\otimes\cH_2\otimes\dots\otimes \cH_N$ to itself.  
We say that a quantum channel is separable if 
all separable states remain separable under the action of this channel. 
We consider a further restricted class of separable channels 
such that all product states remain product states. 
We call these channels {\it completely separable} meaning that 
they do not create any classical correlation. 
Mathematically, a completely separable quantum channel $\Gamma^{sep}$ satisfies 
the condition: 
For all possible states $\rho^{(j)}\in{\cal S}(\cH_i)$, there exist 
some output states $\sigma^{(j)}\in{\cal S}(\cH_i)$ such that 
\be
\Gamma^{sep}(\rho^{(1)}\otimes\rho^{(2)}\otimes \dots\otimes\rho^{(N)})
=\sigma^{(1)}\otimes\sigma^{(2)}\otimes \dots\otimes\sigma^{(N)}, 
\ee
holds. 

When considering completely separable channels, we have the following theorem: 
\begin{theorem}\label{thm4}
Consider a completely separable channel $\Gamma_\theta:\ {\cal S}(\cH^N)\to{\cal S}(\cH^N)$ 
parametrized by a single parameter $\theta$.  
Let $g_\theta^*(\Gamma_\theta)$ be the largest value 
of SLD quantum Fisher information defined by 
\be
g_\theta^*(\Gamma_\theta):=\max_{i}\max_{\rho^{(i)}\in{\cal S}(\cH_i)} 
\underset{j\neq i}{\mathrm{Tr}}\{ \Gamma_\theta(\rho_{cm}^{(1)}\otimes\dots\otimes\rho^{(i)}\otimes\dots\otimes\rho_{cm}^{(N)}  ) \}, 
\ee
with $\rho_{cm}^{(i)}$ the completely mixed state on $\cH_i$. 
For each value $\theta$, 
if a density operator $\rho$ on ${\cal S}(\cH^N)$ is separable, then the SLD quantum Fisher information 
$g_\theta[\Gamma_\theta(\rho)]$ is smaller or equal to the value $N g_\theta^*(\Gamma_\theta) $. 
\end{theorem}
Proof of this theorem goes exactly in the same as Theorem \ref{thm1} as follows. 
\begin{proof}
Consider an arbitrary separable states on ${\cal S}(\cH^N)$ of the form
\be
\rho_{sep}=\sum_{i}p_j\rho_i^{(1)}\otimes\rho_i^{(2)}\otimes \dots\otimes\rho_i^{(N)}. 
\ee
Then, we have the following inequalities. 
\begin{align*}
g_\theta[\Gamma_\theta(\rho_{sep})]&=
g_\theta[\sum_i p_i\Gamma_\theta (\rho_i^{(1)}\otimes \dots\otimes\rho_i^{(N)})] \\ %\label{conv2}
&=g_\theta[\sum_{i} p_i  \sigma_{\theta,i}^{(1)}(\theta)\otimes \dots\otimes\rho_{\theta,i}^{(N)}(\theta)] \\
&\le \sum_{i} p_i g_\theta[\sigma_{\theta,i}^{(1)}(\theta)\otimes \dots\otimes\rho_{\theta,i}^{(N)}(\theta)] \\
&= \sum_{i} p_i  \sum_k g_\theta[\sigma_{\theta,i}^{(k)}(\theta)] \\
&\le \sum_{i} p_i  \sum_k g^*_\theta[\Gamma_\theta] \\
&=Ng^*_\theta[\Gamma_\theta].
\end{align*}
The first inequality follows from the convexity of SLD quantum Fisher information. 
The second inequality follows from the definition of $g^*_\theta[\Gamma_\theta]$ 
and the identification $g_\theta^*(\Gamma_\theta)=\max_{k} \max_{\rho^{(i)}} g_\theta[\sigma_{\theta,i}^{(k)}(\theta)]$. 
$\square$
\end{proof}

To applicate usefulness of Theorem \ref{thm4}, let us consider 
an open quantum system of $N$ qubits described by the following master equation: 
\be\label{mq1}
\frac{\del}{\del t}\rho(t)= \I [\rho(t)\,,\, H_\theta]-\frac14\sum_{i=1}^N\sum_{j=1}^3\gamma_j[[\rho(t)\,,\,\sigma_j^{(i)}]\,,\,\sigma^{(i)}_j],
\ee
where $H_\theta=\frac12\sum_j \theta \sigma^{(i)}_3$ is the free Hamiltonian 
describing a global rotation about an angle $\theta$, 
$\gamma_j$ are the damping parameters that may depend on the parameter of interest $\theta$, 
and $\sigma^{(i)}_j=I\otimes\dots\otimes \sigma_j\otimes\dots\otimes I$ is 
the $j$th Pauli matrix for the $i$th qubit system. 
This kind of master equations has been investigated by several authors 
under the name of noisy quantum metrology, see for example Ref.~\cite{cbmka13}. 
It is straightforward to see that the solution to this master equation is regarded 
as a completely separable channel for a given initial state. 
Thus, we can apply Theorem \ref{thm4} to detect entanglement even in the presence of 
quantum noises described by the above master equation. 
The quantity $g^*_\theta[\Gamma_\theta]$ for this channel is calculated by
\be
g^*_\theta[\Gamma_\theta]=\max_{\rho_0\in\sofc2} g_\theta[ \rho_\theta(t)],
\ee
where $\rho_\theta(t)$ is the solution to the master equation for the single qubit system:
\be\label{mq2}
\frac{\del}{\del t}\rho_\theta(t)=\I [\rho_\theta(t)\,,\, \frac12\theta\sigma_3]-\frac14\sum_{j=1}^3\gamma_j[[\rho_\theta(t)\,,\,\sigma_j]\,,\,\sigma_j],
\ee
with the initial state $\rho(t=0)=\rho_0$. 

The master equation \eqref{mq2} can be solved analytically, but the resulting SLD quantum Fisher 
information gets complicated in general, in particular when the damping coefficients depend on $\theta$. 
Below we consider an isotropic noise $\gamma_1=\gamma_2=\gamma_3=\gamma$ and 
$\gamma$ is independent of $\theta$ to simplify the result. 
In this case, the obtained maximum SLD quantum Fisher information over all possible initial state is 
\be
g^*_\theta[\Gamma_\theta]=t^2 \Exp{-2\gamma t}, 
\ee
at some later time $t$. Thus, the proposed criterion to detect entangled states is as follows. 
For a given initial state $\rho_0$ on $\sofhN$ with $\cH=\bbc^2$,  
$\rho_0$ is entangled if the inequality
\be\label{criterionop1}
g_\theta[\rho_\theta(t)] >N t^2\Exp{-2\gamma t},
\ee
holds for later time $t$. Here $\rho_\theta(t)$ is the solution to the master equation \eqref{mq1} 
with the initial state $\rho_0$. Here, two remarks on this result. 
First, the above criterion seems counterintuitive at first sight. 
Since the right hand becomes exponentially small for fixed $N$ as the time $t$ increases, 
this criterion states that almost all states having non-zero SLD quantum Fisher information 
at later time are entangled. A simple explanation for this observation 
is that as time grows the solution to the master equation \eqref{mq2} approaches 
to $\theta$-independent state, typically to the completely mixed state, for any initial state. 
It is then clear that the amount of SLD quantum Fisher information decreases in time as well. 
Therefore, the inequality \eqref{criterionop1} still provides useful information 
to detect entangled states. 

Second remark is that the above criterion can be weaken by 
replacing $\exp(-2\gamma t)$ by $1$. That is, if the simplified inequality 
\be\label{criterionop2}
g_\theta[\rho_\theta(t)] >N t^2,
\ee
holds, then the state $\rho_0$ is entangled. 
This later criterion \eqref{criterionop2} is certainly simple, 
in particular, it is independent of the external noise parameter $\gamma$. 
However, it is obvious that this weaker version becomes useless for the large $t$ regime. 

In the experiment reported in Ref.~\cite{smlzhpso14}, 
authors apply the weaker version of entanglement criterion 
even though non-negligible decoherence effects are present.  
The above simple example implies that a more sharpened 
criterion can be applied to their experimental data 
by analyzing the effects of quantum noises and to detect 
entangled states faithfully. 

\section{Conclusion} \label{sec6}
We have derived a general criterion to detect entanglement 
based on the SLD quantum Fisher information for any one-parameter family of quantum channels. 
This criterion includes previously known criteria based on unitary channels as a special case. 
We then apply our criterion to detect entanglement in the Bell-diagonal 
states based on the unitary channel, depolarizing channel, and transpose channel. 
Our result shows that even the depolarizing channel 
can be used to detect entangled states for a certain parameter range. 
To put it differently, entanglement is still useful to lower the estimation errors 
even though channels cannot be estimated with quantum metrological enhancement. 
Lastly, we have derived a more general criterion 
that can be applied to the estimation of channel parameter in open quantum systems. 
We briefly discussed how to apply it for the phase estimation in presence 
of a coupling to an environment, which is described by some master equation. 
A more detail discussion on the entanglement detection in open quantum systems 
deserves further studies and it will be analyzed in due course. 

\begin{acknowledgments}
The author wishes to thank M.~Hayashi for bringing his attention 
to Refs.~\cite{hayashi11,hayashi12} and for explaining the details of the papers. 
%He also acknowledges discussion with participants 
%during the 14th asian conference on quantum information science (AQIS14) 
%where a part of results of this paper was presented. 
\end{acknowledgments}


\begin{thebibliography}{99}
\bibitem{comment1}
There are examples where entanglement is not necessary to 
attain the Heisenberg limit if one considers different figure of merit or 
when one concerns the optical quantum metrology, 
i.e., the system with $n$ photon states \cite{thmn10}. 
Here we will not consider such cases.
\bibitem{comment2}
As Hayashi \cite{hayashi11} pointed out, the condition of locally unbiased estimators 
becomes weak when a channel can be estimated quantum metrologically. 
In this case, one should avoid artificial estimators and one possible solution 
was to look for locally asymptotic min-max estimators. 
See also discussion in Ref.~\cite{hayashi12}.
\bibitem{comment3}
The parameter can also be extended to $\Theta=(-1/3,1)$ where 
the channel is still CP-TP, yet we see this extended region is not useful 
to detect entanglement.
\end{thebibliography}
\end{document}